\documentclass[review]{elsarticle}

\usepackage{times}
\usepackage{graphicx}
\usepackage{soul}

\begin{document} 
\begin{frontmatter}
\title{A Robust Opinion Spam Detection Method Against Malicious Attackers in Social Media} 

\author[QOMaddress]{Amir Jalaly Bidgoly\corref{corauth1}}
\ead{jalaly@qom.ac.ir}

\author[QOMaddress]{Zoleikha Rahmanian}

\address[QOMaddress]{Department of Information Technology and Computer Engineering, University of Qom, Qom, Iran.}




\cortext[corauth1]{Corresponding author:}


\begin{abstract}
 Online reviews are potent sources for industry owners and buyers, however, opportunistic people may try to destruct or promote their desired product by publishing fake comments named spam opinion. So far, many models have been developed to detect spam opinions, but none have addressed the issue of spam attacks. It is a way a smart spammer can deceive the system in a manner in which he can continue generating spams without the fear of being detected and blocked by the system. In this paper, the spam attacks are discussed. Moreover, a robust graph-based spam detection method is proposed. The method respectively estimates honesty, trust and reliability values of reviews, reviewers, and products considering possible deception scenarios. The paper also presents the efficiency of the proposed method as compared to other graph-based methods through some case studies.
\end{abstract}

\begin{keyword}
Spam Attack, Spam Detection, Spam Opinion, Deception, Robustness. 
\end{keyword}

\end{frontmatter}

\section{Introduction}

Due to the importance of users' reviews on social networks and commercial websites, many malicious people are struggling to meet their false goals by publishing false opinions. Some of them are trying to promote their non quality goods; others are trying to destroy the quality of their competitors. Thus, researchers have been working since 2007 to provide an appropriate solution to identify spam reviews \citep {RN32}.

In spam detection systems, deception means those so called spam attackers who are aware of the existence of the spam detection method and its mechanism, mislead the system and bypass the mechanism to spread opinion spams according to its malicious goals. For example, in a repeat-based method (e.g.\citep{RN3}), if spam attackers are aware of the system's method, they can easily deceive the system by collecting data from unique and non repetitive reviews. As a further instance, in text-based systems \citep{RN43,RN2}, if the spam attackers are aware of what the words are as a benchmark for spam recognition, they can deceive the system by avoiding their choice.

\begin{figure}
\centering
\includegraphics[width=.6\textwidth]{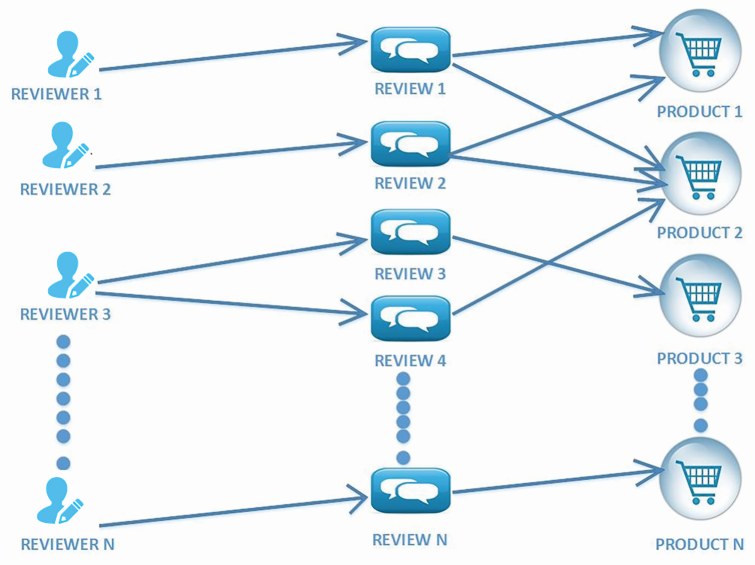}
\caption{The interact of different graph nodes}
\label{fig1}
\end{figure}

 Robustness in these methods is the ability of the system against these deceptive behaviors. In a robust spam detection system, the spam attacker should not be able to deceive the system even with the complete awareness of the detection mechanism and performing deceiving behaviors.  To our knowledge, no previous research has been pointed out the robustness issue of spam detection systems against attacks, and it can be argued that this area of research is still in the early stages.
As can be seen in Fig.\ref{fig1}, this paper proposed an unsupervised graph-based method in which the graph nodes include reviews, reviewers, and products. The proposed method may detect spam attackers in different scenarios. 

For this aim, it calculates the trust score for the reviewer, the honesty score for reviews, and the reliability score for products; their values are updated in an iterative algorithm. The reliability score should represent the true reputation value of the products ignoring spammers. The spams and spam attackers are expected to be identified with having low honesty and trust values.
The proposed method is compared with two other graph-based methods \citep{RN13,RN25}; the results show that the proposed method has improved the system robustness. One challenge to have a complete assessment in spam detection methods is the lack of an appropriate technique to generate spam reviews; thus, the paper also developed a simulation tool, which can produce a large number of reviews tailored to the desired behavior pattern. 

The paper continues as follows. In the next section, related works are reviewed. In section III the motivation and the foundations of the research are presented. The simulator is described in section IV. Further, the proposed method and algorithm are presented in section V and section VI respectively. Section VII describes the method using some case studies, and finally, the paper ends in section VIII with the conclusion and future works.

\section{Related work}

Approaches for detecting spam comments are categorized in several ways. Some divide them into two general types: supervised \citep{RN9,RN8,RN11,RN43,RN2,RN33} and unsupervised methods\citep{RN12,RN25, RN10,RN36}. However, in some other researches, they are categorized into three categories: spam reviews, spam attacker, group spam attacker \citep{RN17}. 

In the set of spam reviews, the contents of reviews are studied and evaluated. One of the most important approaches in this category is repeated-based detection. These methods try to identify spam reviews by going through repetitive patterns of reviews from the same or different reviewers about similar or different products \citep{RN3,RN8}. In addition to this, concept repetition can also be introduced as a measurement criterion for spam comments; the method provided by Alger et al. to identify spam comments \citep{RN20}. Doing multiple counterfeit reviews is time consuming and costly. Spam attackers often do not produce a large number of exclusive counterfeit reviews. They tend to copy the existing text. Therefore, identifying similar opinions is a central part of detecting spam comments. Some literature uses a linguistic character in the text of review \citep{RN28,RN23}.

In the field of identifying spammers, some methods \citep{RN12,RN25,RN36} use inter relationships between the review, the reviewer, and product graph, as shown in Fig.\ref{fig1}, to identify the spammer and also compute the trustworthiness of the reviewer, the honesty of review, and the reliability of the store. 

Some researchers believe that spammers use a specific period to generate spam comments. The numbers of reviews rise dramatically in that interval; thus, they use burst patterns to identify spam attacker\citep{RN27,RN26} .  Some of them, for example, use time series to identify spammers\citep{RN1,RN37}. 
Few pieces of research have been yet done in spammer groups’ identification \citep{RN14,RN47}. The proposed algorithm in the research includes two steps: 1) the extraction of the repeated pattern, and 2) ranking the group based on the spam group indices.

\section{Motivation and research foundations}

Despite the many types of research that have been taken to identify spam comments, so far, no one has been addressing the issue of spam attacks. Today, smart spammers, with the knowledge of spam detection methods, can easily deceive the spam detection system and continue their malicious activity. For example, in text-based systems, Spam attacker deceives the method by modifying the text of reviews.
\begin{figure} 
\centering 
\includegraphics[width=.9\textwidth]{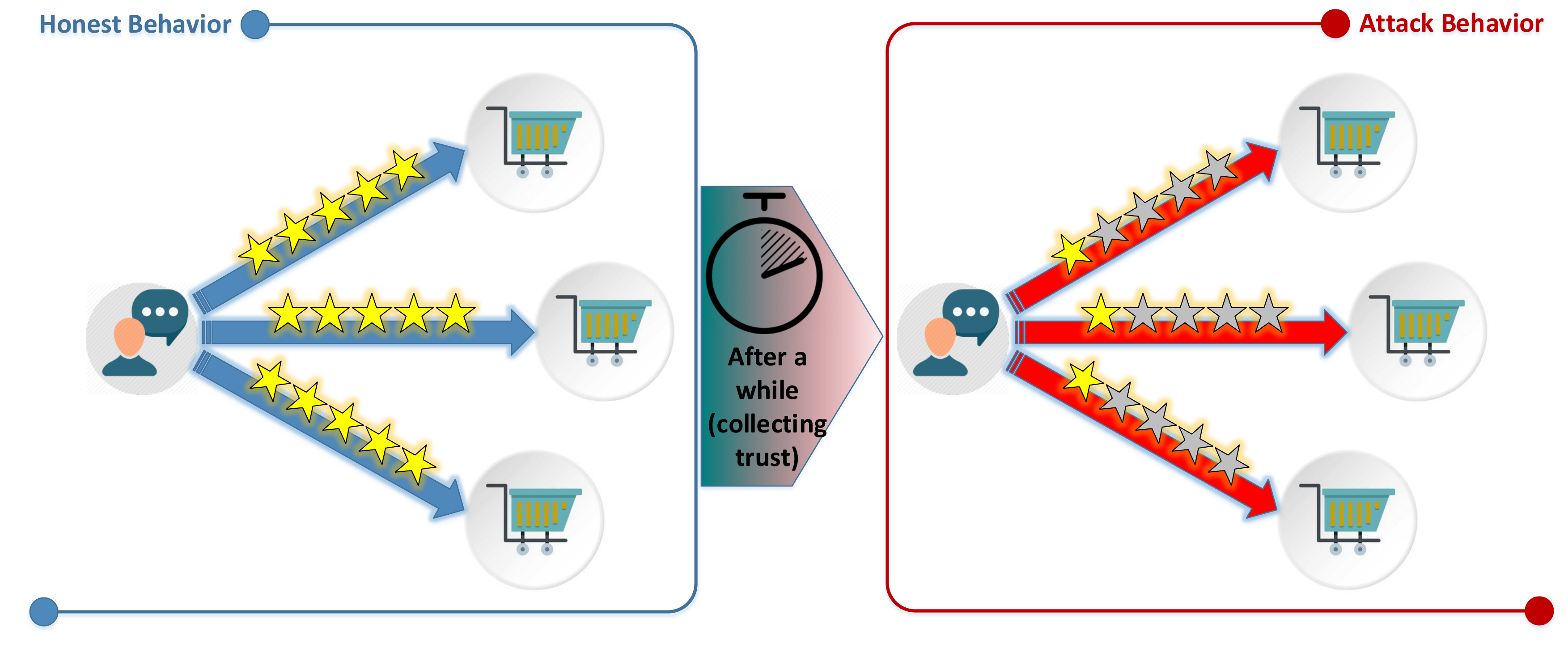}
\caption{User deceptive behaviour over time }
\label{fig2}
\end{figure}

Deception can be performed in two manners: 1) deceptive behavior over time (in length deception), and 2) deceptive behavior over the product (in width deception). In deceptive behavior over time as can be observed in Fig.\ref{fig2} (the quality of all products is 5), spam attacker exhibits honest behavior for a while and after gaining enough trust, discloses its deceitful behavior. It means that smart spammer has a conflicting behavior over time:  honest behavior in the first period to increase his/her credential, and dishonest behavior in the next period to achieve his/her malicious goals using the gained social trust.  

\begin{figure} 
\centering 
\includegraphics[width=.5\textwidth]{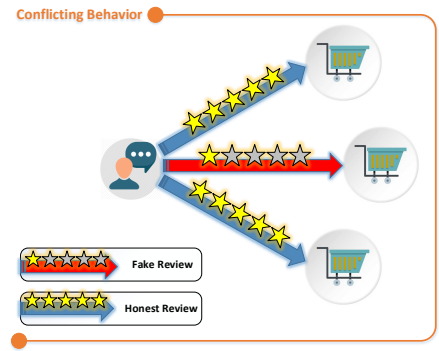}
\caption{User deceptive behavior in width}
\label{fig3}
\end{figure}

In contrast to in length deception, in deceptive behavior over the product (in width deception), the spam attacker exhibits conflicting behavior over different products. As can be seen an instance of this attack in Fig.\ref{fig3}, he sends fake reviews for the product that wants to slander (here's product 2), while writing honest reviews on other products (here's product 2, 3) to keep its social trust value. 

As mentioned earlier, detecting these types of attacks need analyzing the complete knowledge of nodes behaviors, which are used by graph-based spam detection methods; however, the current graph-based techniques almost can be deceived via mentioned attacks.
\section{Simulator software}

Since none of the current spam data sets include the spam attacks, and generation of enough human opinions and performing mentioned attacks are hard in practice, a Simulation tool has been developed on the basis of  PDETool platform\citep{RN52,RN51} to simulate the spam attack scenarios and evaluate the proposed method. The tool simulates the reviewing process in an e-Commerce website and can generate enough samples for any given scenario. Moreover, it is capable of simulating any other desired scenario that may be required in evaluating spam detection methods. 

\begin{figure} 
\centering 
\includegraphics[width=.5\textwidth]{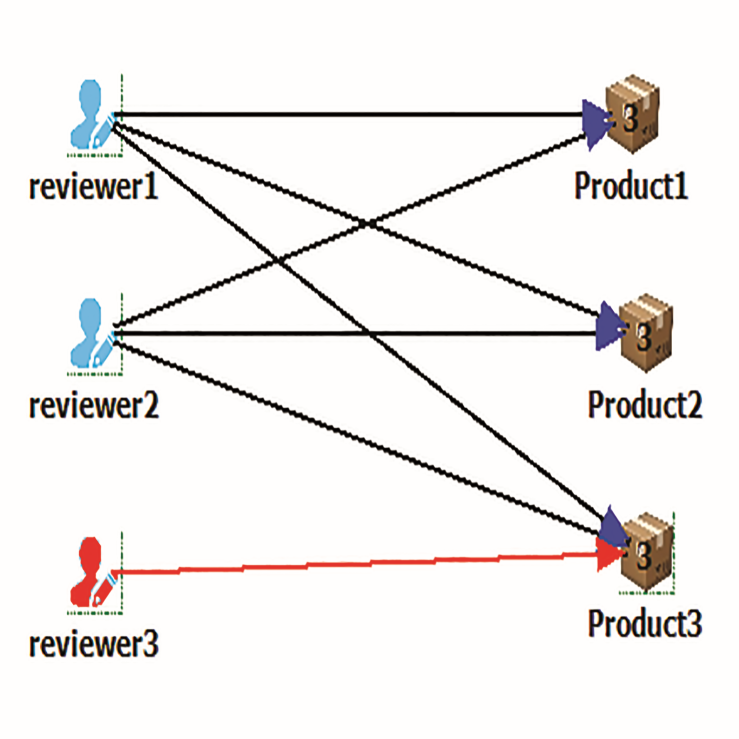}
\caption{A view of the simulator environment}
\label{fig4}
\end{figure}

The tool defines the reviewing environment as a graph, which includes two types of node: 1) product nodes, and 2) reviewer one. Each product has a defined quality. A reviewer should be connected to a product using connectors to produce a review scenario. The reviewer nodes have two sub types: honest and spammer reviewers, which are represented by blue and red users in the tool graphical user interface. The honest reviewers honestly score the products. To be more specific, their score has a normal distribution with the mean of product quality and a given variance. The variance default value is 0.5; however, can be changed by the user.  In contrast, the spammer behavior is defined using a provided script, which enables the modeler to define any complicated scenario including spam attacks. Moreover, the software is also capable of defining individual spam behavior for each product. An instance of the defined mode in this tool is given in Fig.\ref{fig4} As shown in the figure, there are 3 reviewers and 3 products in this model. The model includes a spam reviewer (the 3rd reviewer) who falsely scores the last product (i.e., product no 3) across the red connection.
\section{Proposed model for detecting Spam attacker}

In this section, a method for spam and spam attack detection is proposed, which is robust against mentioned deception scenarios in previous sections. The method is a graph-based model that is defined by three types of nodes: review, reviewer, and product. It estimates the value of trustworthiness, honesty, and reliability for these nodes respectively, which are demarcated in the following in detail. Symbols used in the formulas are given in Table.\ref{table1}.

\begin{table}[]
\caption{ Symbols used in proposed model}
\begin{center}
\label{table1}
\begin{tabular}{|l|l|}
\hline
Definition                                                                                                                                 & Notation      \\ \hline
Review                                                                                                                                     & v             \\ \hline
Reviewer                                                                                                                                   & R             \\ \hline
Product                                                                                                                                    & P             \\ \hline
Score of review v                                                                                                                          & v.score       \\ \hline
Product of review v                                                                                                                        & v.product     \\ \hline
Review of reviewer r                                                                                                                       & r.review      \\ \hline
Reviewer of product p                                                                                                                      & p.reviewer    \\ \hline
\begin{tabular}[c]{@{}l@{}}Reviews of product p \\ by reviewer r\end{tabular}                                                              & p.r.reviews   \\ \hline
\begin{tabular}[c]{@{}l@{}}The maximum difference\\  between the\\  score of reviewer\\  and the majority \\ of the community\end{tabular} & W             \\ \hline
Number of review                                                                                                                           & review.number \\ \hline
\end{tabular}
\end{center}
\end{table}

\textbf{Reviewer Trustworthiness}:
The reviewer's trust score (denoted by $T(r)$ ) is the normal honest behavior performed by the user. It is estimated using the honesty mean of his published reviews and its sequence. The sequence helps the model to give more weight to recent reviews, which is essential for in length spam attack detection. The trust of a reviewer is calculated through the following formula:

\begin{equation}
T(r) = \frac{{\sum\nolimits_{\forall review \in r.review} {review.number*H(r.review)} }}{{\sum\nolimits_{\forall review \in r.review} {review.number} }}
\end{equation}

\textbf{Review Honesty}:
The review's honesty (denoted by $H (v)$ ) indicates the accuracy of the opinion. The honesty value is estimated based on its maximum distant from the estimation of the true quality of the product (i.e., product reliability). The honesty score is defined as follows:
\begin{equation}
\begin{array}{l}
H(v) = \frac{{1 - |normalized(v.score) - R(v.product)|}}{W}\\
\\

R(v.product) > 0.5 \to W = R\\
\\
R(v.product) < 0.5 \to W = 1 - R
\end{array}
\end{equation}

The honesty score is a value between zero and one. The higher value indicates a more honest review. The value of one means that the review is perfectly honest since it fully matches the product reliability. It is not able that the review scores (i.e., $v.score$ should be normalized in the range [0, 1] before being used in the above equation (in the case of the systems uses 1-5 scores).

\textbf{Product Reliability}:
The reliability score of the product (denoted as $R(p)$) is the estimation of the true quality of the product. It depends on both the trust score of the reviewers and their review’s honesty. The product reliability score is defined as follows. The score is a value in the range of [0, 1].
\begin{equation}
R(p) = \frac{{\sum\nolimits_{r \in p.reviewers} {\sum\nolimits_{v \in p.r.reviews} {T(r) * H(v) * v.score} } }}{{\sum\nolimits_{r \in p.reviewers} {\sum\nolimits_{v \in p.r.reviews} {T(r) * H(v)} } }}
\end{equation}

\textbf{Iterative Computation Algorithm}

\begin{figure} 
\centering 
\includegraphics[width=.5\textwidth]{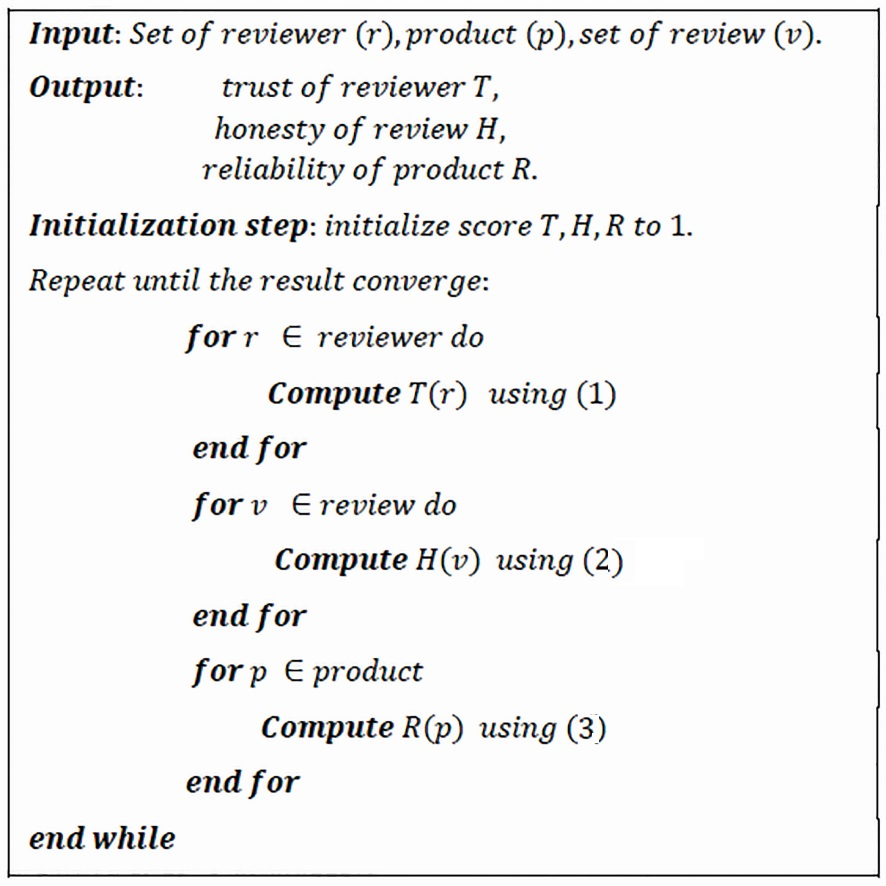}
\caption{The algorithm of the proposed model}
\label{fig5}
\end{figure}

As all trustworthiness, honesty, and reliability values are interdependent for estimating them, the mentioned formulas should be computed in a loop until the result converges to a value. The algorithm output is independent of the initial values of the nodes (trustworthiness, honesty, and reliability). The proposed algorithm for this method can be seen in  Fig.\ref{fig5}.

\section{Deception scenarios and algorithms implementation}

In this section, the efficiency of the proposed method (ROSD ) is presented using some spam attack scenarios. Moreover, the results are compared with other well known graph-based approaches including Wang's \citep{RN36} and Fayazbakhsh's models \citep{RN25}. In all scenarios, 1000 reviews are generated using the simulation tool, and the results of all three methods are presented and compared. Since the result values for ROSD is [0, 1], WNG is [-1, +1], and FYZ is [0, 1], the bench marking is done through the following defined measure: 1) the ability to detect spams and spammers, and 2) the number of deviations that the spam attacker can create in the actual value of product’s reliability. Whatever spam attacker cannot deflect the reliability score of a product from its actual score indicates a better system performance. It is noteworthy to consider Fayazbakhsh's model that calculates only a suspicious score of reviewers and products while has no solution for calculating the suspicious score of reviews. 

\subsection{Simple spamming against a product}

\begin{figure} 
\centering 
\includegraphics[width=.5\textwidth]{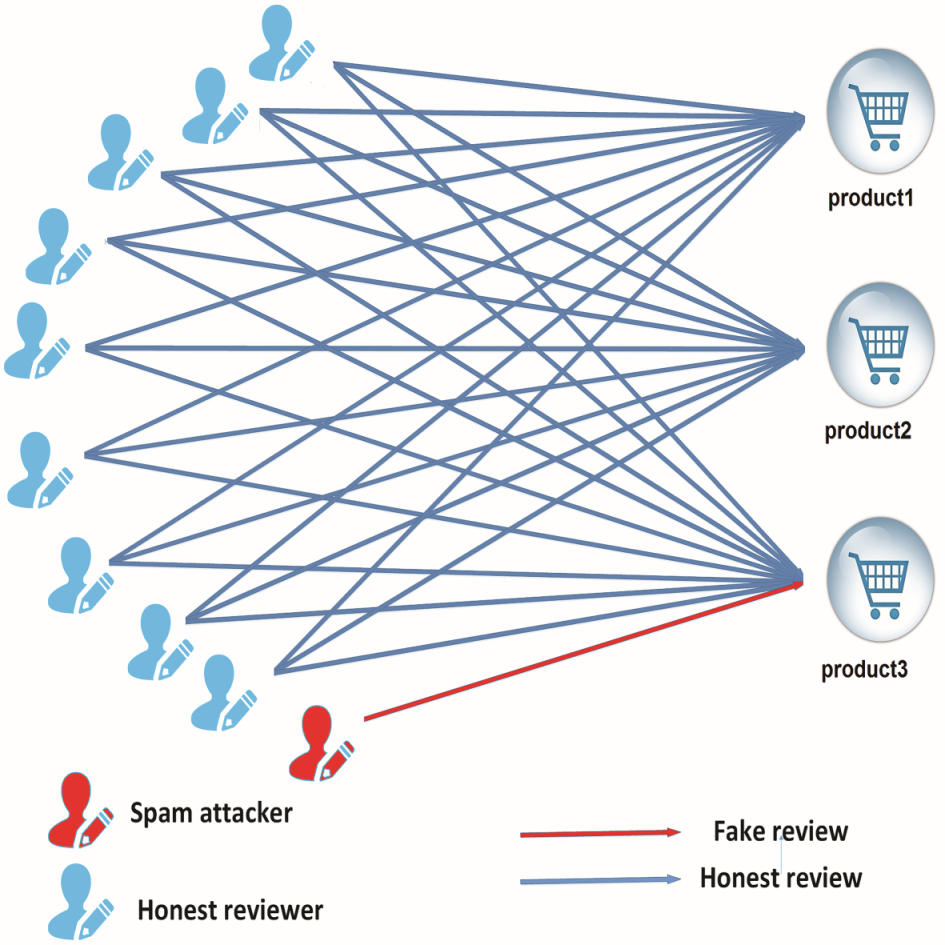}
\caption{Scenario 1: Simple spamming against a product}
\label{fig7}
\end{figure}

In the first scenario, the spammer tries to slander a product without any deception behavior. As can be seen in Fig.\ref{fig7}, there are 10 reviewers and three products. The last reviewer is a spammer who gives zero to product3. In this scenario, spam attacker wants to slander the product and does not use deception scenario. It is important to note that the true quality of all products is considered to be 3 out of 5.

The results are shown in  Table.\ref{table2} As it is presented, ROSD and WNG can find a spammer, while FYZ is unable to detect the spammer as FYZ only tries to find spammer that sends high score (4 or 5 scores). Also in finding the spam reviews, both of the models have acceptable results. As can be seen in the last row of Table.\ref{table2}, in all three models spammer has not been able to change the reliability score of the target product significantly from the actual quality.  Also as shown in Table.\ref{table3}., if spammer simply publishes positive and fake opinion to promote a product in the same conditions and the spammer constantly gives the score 5 out of 5 to the corresponding product then the same results will be achieved. Note that the true quality of the corresponding product (product3) is considered to be 1 out of 5.

\begin{table}[]
\caption{ Results table for the scenario1- slandering a product}
\begin{center}
\label{table2}
\begin{tabular}{|l|l|l|l|}
\hline
Item                                                      & ROSD   & WNG    & FYZ    \\ \hline
The average trustworthiness rating of honest reviewer          & 0.8667 & 1      & 0.9975 \\ \hline
The average trustworthiness rating of spam attacker            & 0      & -1     & 0.9916 \\ \hline
The average honesty  score of  non spam reviews           & 0.8662 & 0.9238 & -      \\ \hline
The average honesty  score of  spam reviews               & 0      & -1     & -      \\ \hline
The average reliability of  target products before spammer & 0.6    & 1      & 0.9942 \\ \hline
The average reliability of  target products after spammer & 0.6067 & 1      & 0.9953 \\ \hline
Deviation value in product reliability                    & 0.006  & 0      & 0.001  \\ \hline
\end{tabular}
\end{center}
\end{table}
\begin{table}[]
\caption{ Results table for the scenario1- promoting a product}
\begin{center}
\label{table3}
\begin{tabular}{|l|l|l|l|}
\hline
Item                                                      & ROSD   & WNG    & FYZ    \\ \hline
The average trustworthiness rating of honest reviewer          & 0.8789 & 1      & 0.9963 \\ \hline
The average trustworthiness rating of spam attacker            & 0      & -1     & 0.9967 \\ \hline
The average honesty  score of  non-spam reviews           & 0.8759 & 0.8917 & -      \\ \hline
The average honesty  score of  spam reviews               & 0      & -1     & -      \\ \hline
The average reliability of  target products before spammer & 0.2    & -1     & 0.9941 \\ \hline
The average reliability of  target products after spammer & 0.1915 & -1     & 0.9954 \\ \hline
Deviation value in product reliability                    & 0.0085 & 0      & 0.001  \\ \hline
\end{tabular}
\end{center}
\end{table}

\subsection{ An over product attack }

\begin{figure} 
\centering 
\includegraphics[width=.5\textwidth]{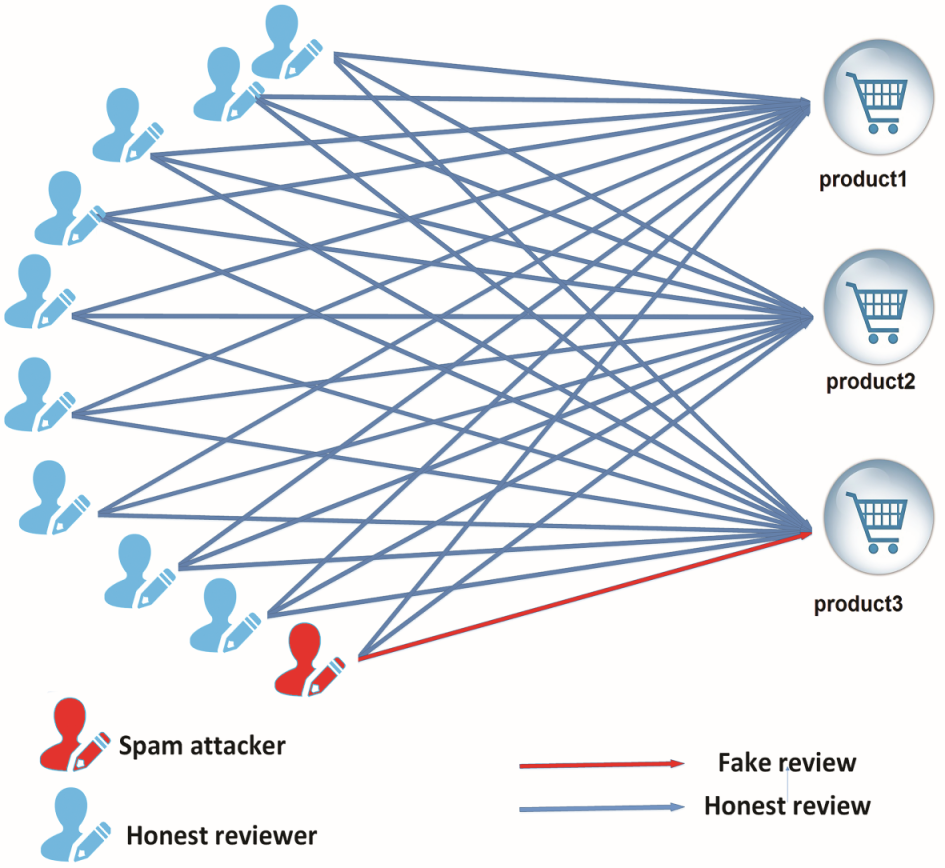}
\caption{Scenario 2: An over product attack}
\label{fig8}
\end{figure}

In the second scenario, an over product attack is simulated. The simulation model is represented in Fig.\ref{fig8}. 8 and is similar to previous ones; however, the spammer is connected to all products. He gives a right score to all products but the last one.
As it is presented, only the proposed model can find spam attackers, and this suggests that other models are deceived in this way since they are unable to find the spam attacker. However, in finding the spam review, both models have acceptable results. As can be seen in the last row of Table.\ref{table4}, in all three models, spam attacker has not been able to change the reliability score of the target product critically from the actual quality. Also, in this scenario, if spam attacker tries to promote a product and using false positive scores instead of slandering, as indicated in Table.\ref{table5}, the same results have been achieved (Here the true score of the selected product is assumed to be 1 out of 5).
\begin{table}[]
\caption{ Results table for the scenario 2- Selective product slandering}
\begin{center}
\label{table4}
\begin{tabular}{|l|l|l|l|}
\hline
Item                                                      & ROSD   & WNG    & FYZ    \\ \hline
The average trustworthiness rating of honest reviewer          & 0.8719 & 1      & 0.9960 \\ \hline
The average trustworthiness rating of spam attacker            & 0.5730 & 1      & 0.9966 \\ \hline
The average honesty  score of  non spam reviews           & 0.8684 & 0.9171 & -      \\ \hline
The average honesty  score of  spam reviews               & 0      & -1     & -      \\ \hline
The average reliability of  target products before spammer & 0.6    & 1      & 0.9965 \\ \hline
The average reliability of  target products after spammer & 0.6060 & 1      & 0.9952 \\ \hline
Deviation value in product reliability                    & 0.006  & 0      & 0.0013 \\ \hline
\end{tabular}
\end{center}
\end{table}

\begin{table}[]
\caption{Results table for the scenario 2- Selective product promoting}
\begin{center}
\label{table5}
\begin{tabular}{|l|l|l|l|}
\hline
Item                                                      & ROSD   & WNG    & FYZ    \\ \hline
The average trustworthiness  rating of honest reviewer          & 0.8704 & 1      & 0.9961 \\ \hline
The average trustworthiness  rating of spam attacker            & 0.5865 & 1      & 0.9962 \\ \hline
The average honesty  score of  non spam reviews           & 0.8726 & 0.8985 & -      \\ \hline
The average honesty  score of  spam reviews               & 0      & -1     & -      \\ \hline
The average reliability of  target products before spammer & 0.2    & -1     & 0.9935 \\ \hline
The average reliability of  target products after spammer & 0.2016 & -1     & 0.9952 \\ \hline
Deviation value in product reliability                    & 0.0016 & 0      & 0.002  \\ \hline
\end{tabular}
\end{center}
\end{table}

\subsection{ An over time attack}

\begin{figure} 
\centering 
\includegraphics[width=.5\textwidth]{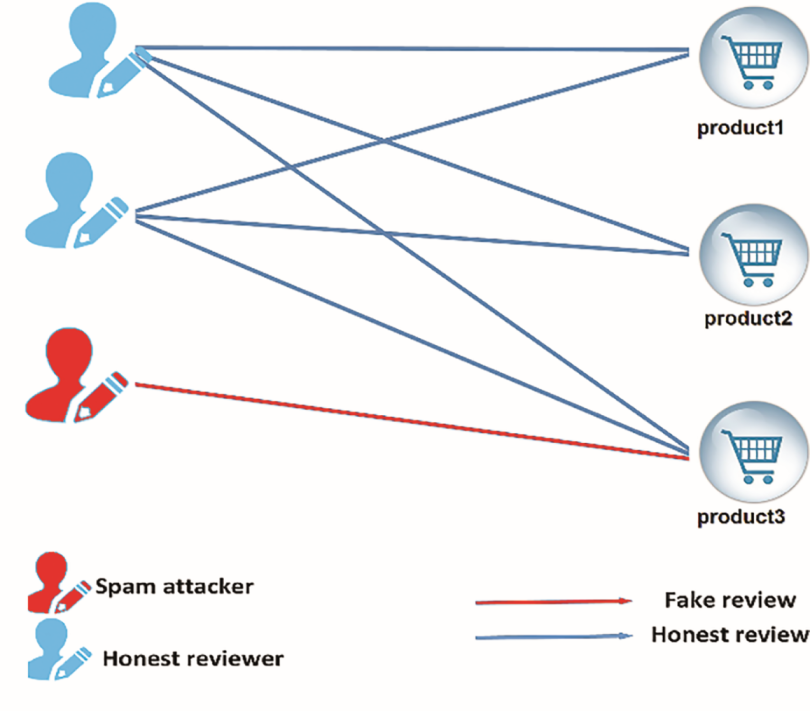}
\caption{Scenario 3: An over time attack}
\label{fig9}
\end{figure}

In this scenario, a slandering attack over time is simulated. As can be seen in Fig.\ref{fig9}, there are 3 reviewers and 3 products. The last reviewer is spam attacker who gives a true score of 3 to product 3 in the intervals of time (20 first reviews), and then gives the score 1 in the intervals of time (20-second reviews); this process continues until the end of the review generation. It should be noted that the true quality of all products is considered to be 3 out of 5.

As illustrated in Table.\ref{table6}, in this scenario results are similar to the earlier scenario, and the proposed model only may detect spam attacker. The results for promoting attack over time are the same; it is shown in Table.\ref{table7}.

\begin{table}[]
\caption{Results table for the scenario3- slandering attack over time}
\begin{center}
\label{table6}
\begin{tabular}{|l|l|l|l|}
\hline
Item                                                      & ROSD   & WNG    & FYZ    \\ \hline
The average trustworthiness  rating of honest reviewers         & 0.8651 & 1      & 0.9814 \\ \hline
The average trustworthiness  rating of spam attacker            & 0.5285 & 1      & 0.9890 \\ \hline
The average honesty  score of  non spam reviews           & 0.9081 & 0.8694 & -      \\ \hline
The average honesty  score of  spam reviews               & 0.3486 & -1     & -      \\ \hline
The average reliability of  target products before spammer & 0.6    & 1      & 1      \\ \hline
The average reliability of  target products after spammer & 0.5736 & 1      & 0.9816 \\ \hline
Deviation value in product reliability                    & 0.0264 & 0      & 0.0184 \\ \hline
\end{tabular}
\end{center}
\end{table}

\begin{table}[]
\caption{Results table for the scenario3- promoting attack over time}
\begin{center}
\label{table7}
\begin{tabular}{|l|l|l|l|}
\hline
Item                                                      & ROSD   & WNG    & FYZ    \\ \hline
The average trustworthiness  rating of honest reviewers         & 0.8678 & 1      & 0.9900 \\ \hline
The average trustworthiness  rating of spam attacker            & 0.5570 & 1      & 0.9926 \\ \hline
The average honesty  score of  non spam reviews           & 0.9081 & 0.8694 & -      \\ \hline
The average honesty  score of  spam reviews               & 0.3486 & -1     & -      \\ \hline
The average reliability of  target products before spammer & 0.6    & 1      & 1      \\ \hline
The average reliability of  target products after spammer & 0.6181 & 1      & 0.9868 \\ \hline
Deviation value in product reliability                    & 0.0181 & 0      & 0.0132 \\ \hline
\end{tabular}
\end{center}
\end{table}

\subsection{Selective product slandering with real data}

In this scenario, an over product attack on real data is implemented. 20 spam data records have been artificially added to an existing spam data set to reach this goal \citep{RN50}. 
This data set is the opinions collected from the movie Lenz website and includes 16 reviewers and 670 products. To emulate the deception scenario in this data, a spam attacker is added to the data that gives 0.5 (the lowest score in real data is 0.5) to some goal products and sends the correct score (similar to honest reviews) to other products. 
The average score of the attacker's target products is about 3.75, so his reviews should definitely be identified as spam. 
The results are shown in Table.\ref{table8} As it is presented, only the proposed model can find spam attackers, and in the case of finding the spam review, both models have acceptable results. However, as can be seen in the last row of Table 8, in the proposed model, FYZ spam attacker has not been able to change greatly the reliability score of the target product from the actual quality.

\begin{table}[]
\caption{Results table for the scenario 4}
\begin{center}
\label{table8}
\begin{tabular}{|l|l|l|l|}
\hline
Item                                                      & ROSD   & WNG     & FYZ \\ \hline
The average trustworthiness  rating of honest reviewer          & 0.9103 & 0.9307  & 1   \\ \hline
The average trustworthiness  rating of spam attacker            & 0.5596 & 0.9682  & 1   \\ \hline
The average honesty  score of  non spam reviews           & 0.9140 & 0.3823  & -   \\ \hline
The average honesty  score of  spam reviews               & 0.1167 & -0.1245 & -   \\ \hline
The average reliability of  target products before spammer & 0.9106 & 0.8230  & 1   \\ \hline
The average reliability of  target products after spammer & 0.8604 & 0.2506  & 1   \\ \hline
Deviation value in product reliability                    & 0.0502 & 0.5724  & 0   \\ \hline
\end{tabular}
\end{center}
\end{table}

\subsection{Selective product promoting with real data}

In this scenario, all the condition is the same as the previous scenario; however, to emulate the deception scenario in this data, a spam attacker is added to the data that gives 5 (the highest score in real data is 5) to some goal products and sends the correct score (similar to honest reviews) to other products. The average score of the attacker's target products is about 1.5, so his reviews should definitely be identified as spam. As before only the proposed model performs well Table.\ref{table9}.

\begin{table}[]
\caption{Results table for the scenario 5}
\begin{center}
\label{table9}
\begin{tabular}{|l|l|l|l|}
\hline
Item                                                      & ROSD   & WNG     & FYZ \\ \hline
The average trustworthiness  rating of honest reviewer          & 0.9119 & 0.9307  & 1   \\ \hline
The average trustworthiness  rating of spam attacker            & 0.5015 & 0.9886  & 1   \\ \hline
The average honesty  score of  non spam reviews           & 0.9154 & 0.3844  & -   \\ \hline
The average honesty  score of  spam reviews               & 0      & -0.024  & -   \\ \hline
The average reliability of  target products before spammer & 0.2053 & -0.6484 & 1   \\ \hline
The average reliability of  target products after spammer & 0.2053 & 0.3760  & 1   \\ \hline
Deviation value in product reliability                    & 0      & 1.022   & 0   \\ \hline
\end{tabular}
\end{center}
\end{table}

\section{Conclusion}

Given the growing importance of users' comments in the virtual world, providing a robust method for detecting smart spam attackers is indispensable. In this paper, a model that would be robust against various types of spam deception behaviors has been proposed. The efficiency of the proposed method is studied in some attack scenarios and compared with two famous and well known models in this area. The results show that not only the proposed model can find spam attacker in every deception scenario, but also has a considerable improvement over the other model. It is capable of detecting the spammer and decreasing its trust. Moreover, it does not allow the attacker to deviate the product reputation to his malicious goals. As future work, it could be useful to implement more deceptive scenarios on review spam detection models and resist the currently proposed model against other deceptive scenarios.

\bibliographystyle{apa}
\bibliography{references}
\end{document}